\documentclass[aps,prl,twocolumn,superscriptaddress,showpacs]{revtex4}
\usepackage{amsmath,graphics}
\newcommand{\D}{\overline{\cal D}}

\begin{document}

Phys. Rev. Lett. \textbf{91}, 014101 (2003)

\title{Heterogeneity in oscillator networks: Are smaller worlds easier to synchronize?}


\newcommand{\csse}{\affiliation{Department of Mathematics, 
Arizona State University, Tempe, Arizona 85287}}
\newcommand{\ee}{\affiliation{Department of Electrical Engineering,
Arizona State University, Tempe, Arizona 85287}}

\author{Takashi Nishikawa}
\email[E-mail: ]{tnishi@chaos6.la.asu.edu}
\altaffiliation[Current address: ]{
    Department of Mathematics,
    208 Clements Hall, PO Box 750156,
    Southern Methodist University,
    Dallas, TX 75275-0156.
}
\csse

\author{Adilson E. Motter}
\email[E-mail: ]{motter@mpipks-dresden.mpg.de}
\altaffiliation[Current address: ]{
  Max Planck Institute for the Physics of Complex Systems
  Nothnitzer Strasse 38, 01187 Dresden, Germany.
}
\csse
	
\author{Ying-Cheng Lai}
\csse
\ee

\author{Frank C. Hoppensteadt}
\csse
\ee

\date{\today}


\begin{abstract}
Small-world and scale-free networks are known to be more easily synchronized than regular lattices, which is usually attributed to the smaller network distance between oscillators.  Surprisingly, we find that networks with a homogeneous distribution of connectivity are more synchronizable than heterogeneous ones, even though the average network distance is larger.  We present numerical computations and analytical estimates on synchronizability of the network in terms of its heterogeneity parameters.  Our results suggest that some degree of homogeneity is expected in naturally evolved structures, such as neural networks, where synchronizability is desirable.
\end{abstract}


\maketitle

In their seminal work, Watts and Strogatz~\cite{watts1998,watts1999} have shown that real networks of different nature have the small-world (SW) property, characterized by high clustering and the average network distance between two nodes that is as small as pure random networks.  
Since then, many more examples of real-world networks%
~\cite{albert1999,amaral2000,strogatz2001,newman2001,montoya2002,motter2002}, including both artificial and natural systems, have been identified to have the SW property.  It turns out, however, that there is another seemingly generic feature of networks in the real world.  It is called the scale-free (SF) property, which is signified by the power-law connectivity distribution of the network~\cite{barabasi1999,albert1999,faloutsos1999,amaral2000,newman2001b,liljeros2001,%
sole2001,albert2002,motter2002}.  Barab\'asi and Albert~\cite{barabasi1999} suggested a model of growing networks, in which preferential attachment of new links to nodes with higher connectivity results in the SF property.  There has also been some efforts to incorporate both SF and SW properties in a single model~\cite{amaral2000,newman2001c,klemm2002}.

So far, much research has been focused on the structural properties of SF and SW network models.  Despite the widespread belief that these structural properties must have significant impact on dynamical processes taking place on such networks~\cite{strogatz2001}, there has been little work directly addressing this issue.  Most work has dealt with synchronization of oscillators whose topology of interaction has either the SF or SW%
~\cite{watts1999,lago-fernandez2000,gade2000,jost2001,hong2002,wang2002,barahona2002,kwon2002} property, showing that it leads to improved synchronizability when compared to local lattice topology.  A general argument underlying this phenomenon is that communication between oscillators are more efficient because of the smaller average network distance.  But, does smaller average network distance improve synchronizability?

For many real networks, heterogeneity is a common trait which frequently manifests itself in the form of an SF distribution of connectivities.  It is known that such heterogeneity tends to reduce the average network distance~\cite{CohenBollobas}, and this leads naturally to the question of whether heterogeneity improves synchronizability. 


The aim of this paper is to demonstrate, by using important classes of SF and SW network models, that heterogeneity of the connectivity distribution causes the opposite to hold; namely, as heterogeneity increases, the average network distance is reduced but synchronization becomes more difficult to achieve.  We show that this intriguing behavior can be explained by examining the load distribution on nodes or links, where the load of a node (or a link) quantifies the traffic of communication passing through it.  The analytical results we derive suggest that our observations are quite general and are expected to hold for a wide class of complex networks.

Synchronizability of a network of oscillators can be quantified
through the eigenvalue spectrum of the Laplacian matrix representing
the connection topology of the network.  Here we follow the general
framework established in~\cite{barahona2002,pf}.
Consider a network of $N$ identical dynamical systems with symmetric
coupling between oscillators.  The equations of motion for the system
are
\begin{equation}
\dot{x_i} = F(x_i) + \sigma \sum_{j=1}^N L_{ij} H(x_j),\;\; i = 1,\ldots,N,
\end{equation}
where $\dot{x} = F(x)$ governs the dynamics of each individual node,
$H(x)$ is the output function, $\sigma$ is the overall strength of
coupling, and $L$ is the Laplacian matrix, defined to be $L_{ij} = -1$
if nodes $i$ and $j$ are connected, $L_{ii} = k_i$ if node $i$ is
connected to $k_i$ other nodes, and $L_{ij} = 0$ otherwise.  The
linear stability of the synchronized state $\{ x_i(t) = x^*(t), \forall
i \}$ is determined by the corresponding variational equations, which
can be diagonalized into $N$ blocks of the form
$\dot{y} = [DF(s) + \lambda DH(s)]y$,
where $y$ represents different modes of perturbation from the synchronized state.  We have $\lambda = \sigma\lambda_i$ for the $i$th block, $i = 1,2, \ldots, N$, and $\lambda_1 = 0 \le \lambda_2 \le \ldots \le \lambda_N$ are the eigenvalues of $L$~\cite{footnote1}. The largest Lyapunov exponent $\Lambda(\lambda)$ for this equation, also called the master stability function~\cite{pf}, determines the linear stability of the synchronized state for \emph{any} linear coupling scheme.  In particular, the synchronized state is stable if $\Lambda(\sigma \lambda_i) < 0$, for each $i = 2,\ldots,N$~\cite{footnote2}.  It was found~\cite{pf} that for a large class of chaotic oscillatory systems, there exists a single parameter interval $(\alpha_1, \alpha_2)$ on which $\Lambda(\lambda) < 0$.  In this case, there is a value of the coupling strength $\sigma$ for which the synchronized state is linearly stable, if and only if $ \lambda_N/\lambda_2 < \alpha_2/\alpha_1 \equiv \beta$, where $\beta$ is a constant that depends on $F(x)$, $x^*(t)$, and $H(x)$, but not on $L$.  The value of $\beta$ ranges from 5 to 100 for various chaotic oscillators~\cite{barahona2002}.  Realizing that the ratio $\lambda_N/\lambda_2$ depends only on the topology of interactions among oscillators, we see that the impact of having a particular coupling topology on the network's ability to synchronize is represented in a single quantity $\lambda_N/\lambda_2$: the larger the ratio, the more difficult it is to synchronize the oscillators, and vice versa~\cite{barahona2002}.

Having reduced the problem of synchronizability to finding eigenvalues of the Laplacian matrix $L$, we now study the effect of heterogeneity in the connectivity distribution. We first consider the semirandom model of SF networks~\cite{newman2001c}, where the connectivity $k_i \ge k_0$ of each node $i$ is chosen at random according to the probability distribution $P(k) \sim k^{-\gamma}$ with the scaling exponent $\gamma$.  An SF network is then generated by randomly connecting nodes, forcing each node $i$ to have connectivity $k_i$, and prohibiting self- and repeated links~\cite{newman2001c,motter2002b}.

\begin{figure}
\setlength{\unitlength}{3.3in}
\begin{picture}(1,0.52)
\put(0.99,0.27){\makebox(0,0)[r]{
\resizebox{0.93\unitlength}{!}{
\includegraphics{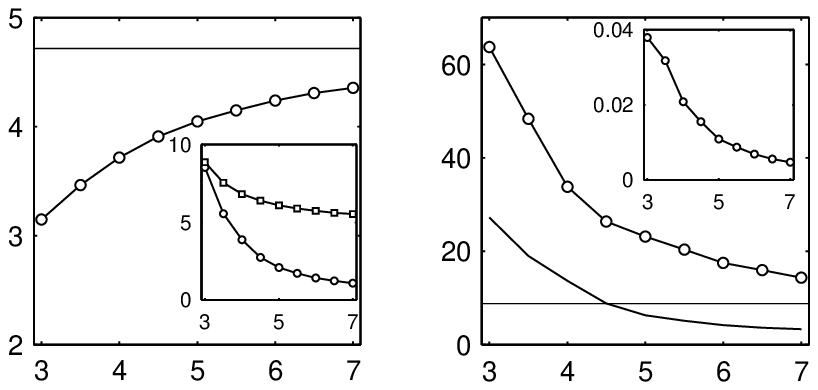}
}}}

\put(0.13,0.49){\makebox{(a)}}
\put(0.64,0.49){\makebox{(b)}}

\put(0.285,0.01){\makebox(0,0)[b]{$\gamma$}}
\put(0.415,0.1){\scalebox{0.7}{\makebox(0,0)[b]{$\gamma$}}}
\put(0.39,0.2){\scalebox{0.7}{\makebox(0,0)[b]{$s$}}}
\put(0.39,0.27){\scalebox{0.7}{\makebox(0,0)[b]{$\bar{k}$}}}
\put(0.02,0.28){\makebox(0,0)[l]{$\D$}}
\put(0.12,0.395){\makebox(0,0)[l]{$\gamma = \infty$}}

\put(0.79,0.01){\makebox(0,0)[b]{$\gamma$}}
\put(0.48,0.28){\makebox(0,0)[l]{$\displaystyle\frac{\lambda_N}{\lambda_2}$}}
\put(0.88,0.39){\scalebox{0.7}{\makebox(0,0){$\ell_{\max}$}}}
\put(0.9,0.24){\scalebox{0.7}{\makebox(0,0){$\gamma$}}}
\put(0.61,0.12){\scalebox{0.7}{\makebox(0,0)[l]{$\gamma = \infty$}}}
\end{picture}

\begin{picture}(1,0.45)
\put(0.985,0.24){\makebox(0,0)[r]{
\resizebox{0.93\unitlength}{!}{
\includegraphics{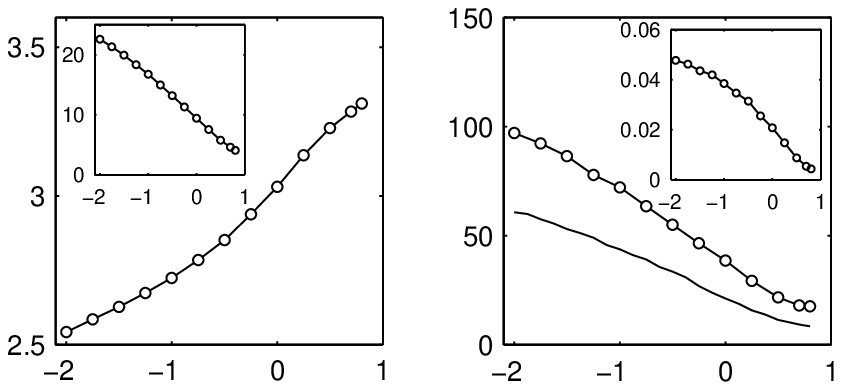}
}}}

\put(0.13,0.45){\makebox{(c)}}
\put(0.64,0.45){\makebox{(d)}}

\put(0.3,0.01){\makebox(0,0)[b]{$\alpha$}}
\put(0.245,0.21){\scalebox{0.7}{\makebox(0,0)[b]{$\alpha$}}}
\put(0.27,0.36){\scalebox{0.8}{\makebox(0,0)[b]{$s$}}}
\put(0.03,0.28){\makebox(0,0)[l]{$\D$}}

\put(0.78,0.01){\makebox(0,0)[b]{$\alpha$}}
\put(0.48,0.25){\makebox(0,0)[l]{$\displaystyle\frac{\lambda_N}{\lambda_2}$}}
\put(0.89,0.37){\scalebox{0.7}{\makebox(0,0){$\ell_{\max}$}}}
\put(0.86,0.2){\scalebox{0.7}{\makebox(0,0){$\alpha$}}}
\end{picture}
\caption{Synchronizability of SF networks of size $N = 2^{10}$.  (a,b) The average network distance $\D$ and the eigenvalue ratio $\lambda_N/\lambda_2$ for the semirandom model with $k_0 = 5$.  The inset of (a) shows the mean $\bar{k}$ and the standard deviation $s$ of the connectivity distribution.  The maximum $\ell_{\max}$ of the normalized load on nodes is plotted in the inset of (b).  The horizontal lines in (a) and (b) indicate the values of $\D$ and $\lambda_N/\lambda_2$ computed for the $\gamma = \infty$ case. (c,d) The same quantities for the growing model with aging of nodes for $n_0 = 5$ and $m_a = 5$.  Note that $\bar{k}$ is not shown because it is trivially $\bar{k} = m_a$.  The solid curves in (b) and (d) are the lower bounds given in Eq.~\eqref{eqn:bd}.  The upper bounds of Eq.~\eqref{eqn:bd} are above the limits, but follow the same trend.  All quantities are averaged over 100 realizations. \label{fig:2}}
\end{figure}

The average network distance $\D$ of an SF network decreases with increasing heterogeneity of the connectivity distribution, where $\D$ is defined as the minimum number of links that must be followed to go from one node to another, averaged over all pairs of nodes.  Figure~\ref{fig:2}(a) shows the dependence of $\D$ on the scaling exponent $\gamma$ for the semirandom model.  In order to quantify the heterogeneity of the connectivity distribution, we plot its standard deviation $s$ in the inset of Fig.~\ref{fig:2}(a).  We observe that smaller values of $\gamma$ result in a longer tail in the connectivity distribution, which in turn makes $s$ larger.  Note that for completely homogeneous case of $\gamma = \infty$, which corresponds to taking $k_i = k_0$ for all $i$, we have $s = 0$.  Note also that decreasing $\gamma$ increases the average connectivity $\bar{k}$ as well, which can be seen in the same inset.  This implies that the total number of connections increases.  In short, the more heterogeneous the network is, the more connected and ``smaller'' it is.

Since the SF and SW networks are known to show enhanced synchronizability over lattices, one might expect that an SF network with smaller $\D$ would have improved synchronizability.  However, we observe the opposite phenomenon: the ratio $\lambda_N/\lambda_2$ increases as $\gamma$ is increased, as shown in Fig.~\ref{fig:2}(b).  In other words, {\it as the network becomes more heterogeneous, it becomes less synchronizable, even though $\bar{k}$ gets larger and $\D$ gets smaller}~\cite{footnote3}.

The heuristic reason for this surprising behavior lies in the fact that a few ``center'' oscillators interacting with a large number of other oscillators tend to get overloaded by the traffic of communication passing through them.  When too many independent signals with
different phases and frequencies are going through a node at the same time, they can cancel one another, resulting in effectively no communication between oscillators.  Hence, the more concentrated the traffic is on a few nodes, the more difficult it is to achieve efficient communication between oscillators, leading to reduced synchronizability.  The same effect results from overloading links as well.

The amount of communication traffic passing through node $i$ can be quantified by the load on $i$, defined as the number of shortest paths between two (other) nodes that pass through $i$~\cite{newman2001b,goh2001}.  To quantify the extent to which the load distribution is concentrated on a few nodes, we first normalize the load on each node by the total load of the network, and then focus our attention on its maximum value $\ell_{\max}$ over all nodes.  The dependence of $\ell_{\max}$ on $\gamma$ is shown in the inset of Fig.~\ref{fig:2}(b).  We see that the behavior of $\ell_{\max}$ is followed closely by the ratio $\lambda_N/\lambda_2$.  This is consistent with the result in~\cite{goh2001} for another model of SF networks.  It is shown there that $\ell_{\max}$ for the largest hub scales as $N^{-\delta}$, where the scaling exponent $\delta \approx 0.2$ for $2 < \gamma \le 3$, but $\delta$ gets larger as $\gamma > 3$ is increased.  This means that for a fixed $N$, more heterogeneous connectivity distribution implies more heterogeneous load distribution on nodes.

To illustrate the prevalence of this phenomenon, we next consider a growing model of SF networks with aging of nodes~\cite{dorogovtsev2000b}.  In this model, the network initially has $n_0$ nodes.  At each time step, a new node with $m_a$ links is added.  The probability for an existing node $i$ to receive a link from the new node is proportional to $k_i$ and $\tau_i^{-\alpha}$, where $\tau_i$ is the age of node $i$, or the number of time steps since it was added to the network.  It was shown in~\cite{dorogovtsev2000b} that the parameter $\alpha$ determines the scaling exponent $\gamma$ for the resulting SF network, and that smaller $\alpha$ $(\le 1)$ yields smaller $\gamma$ $(\ge 2)$.  Figures~\ref{fig:2}(c,d) show the result for the growing model, from which one observes the same phenomenon of reduced synchronizability due to heterogeneity.

To show that any type of heterogeneity induces the same behavior, we now introduce a variant of the SW model~\cite{watts1998}, which is constructed as follows.  First we start with a regular one-dimensional lattice having $N$ nodes with connections up to $\kappa$th nearest neighbors.  We next choose $n_c$ nodes at random from all nodes with equal probabilities, and assign them to be centers, or hubs.  Finally, we add each of $m$ shortcuts by connecting one node chosen at random from all $N$ nodes to another node randomly chosen from the $n_c$ center nodes.  We call this the two-layer SW model since nodes can be naturally divided into two groups: centers, which tend to have higher connectivity, and the others that have lower connectivity.  Thus, the heterogeneity is controlled by the parameter $n_c$: small $n_c$ leads to higher connectivity of the centers, which in turn results in increased heterogeneity. 


Two extreme cases in our model exhibit interesting features. When $n_c = N$, shortcuts are simply added uniformly at random, effectively reducing the network to the homogeneous SW model of Watts and Strogatz~\cite{watts1998,newman2000}.  If $n_c = 1$, all shortcuts are connected to a single center, making the configuration close to the one which minimizes the average network distance $\D$~\cite{dorogovtsev,nishikawa2002}.  

\begin{figure}
\setlength{\unitlength}{3.3in}
\begin{picture}(1,0.52)
\put(0.485,0.27){\makebox(0,0){
\resizebox{\unitlength}{!}{
\includegraphics{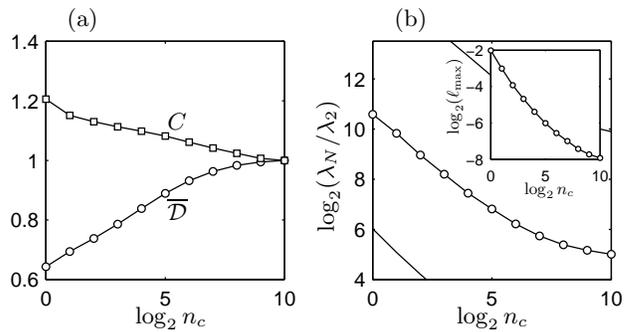}}}}

\put(0.12,0.52){\makebox(0,0)[t]{(a)}}
\put(0.26,0.01){\makebox(0,0)[b]{$\log_2 n_c$}}
\put(0.27,0.34){\makebox(0,0){$C$}}
\put(0.27,0.2){\makebox(0,0){$\D$}}

\put(0.64,0.52){\makebox(0,0)[t]{(b)}}
\put(0.77,0.01){\makebox(0,0)[b]{$\log_2 n_c$}}
\put(0.49,0.27){\makebox(0,0)[l]{\rotatebox{90}{$\log_2(\lambda_N/\lambda_2)$}}}
\put(0.71,0.38){\scalebox{0.7}{\makebox(0,0){\rotatebox{90}{$\log_2 (\ell_{\max})$}}}}
\put(0.86,0.23){\scalebox{0.7}{\makebox(0,0){$\log_2 n_c$}}}
\end{picture}
\caption{(a) The average network distance $\D$, the clustering
coefficient $C$, and (b) the ratio $\lambda_N/\lambda_2$ for the
two-layer model.  The inset in (b) shows the maximum value
$\ell_{\max}$ of the normalized load on nodes.  The solid curves in (b) are the bounds given in Eq.~\eqref{eqn:bd}.  All estimates are the results of averaging over 100 realizations, and $\D$ and $C$ are normalized by the respective values for $n_c = N$: $\D \approx 4.6$ and $C \approx 0.52$.  Other parameters are $N = 2^{10}$, $\kappa = 4$, and $m =
2^9$. \label{fig:1}}
\end{figure}

The network exhibits a stronger SW property when $n_c$ is smaller, i.e., when the network is heterogeneous.  As can be seen in Fig.~\ref{fig:1}(a), decreasing $n_c$ reduces $\D$ and increases the clustering coefficient $C$, where $C$ is defined as the probability that two nodes are connected, given that they are connected to a third common node~\cite{watts1999}.  However, the stronger SW property has a counter-intuitive effect on synchronizability: {\it as the network becomes more heterogeneous (and its SW property stronger), the ratio $\lambda_N/\lambda_2$ increases sharply (up to a factor of about 50), indicating that it becomes more difficult to synchronize the oscillators.}  It is interesting to compare this with the result of~\cite{barahona2002}, in which the network was never observed to become less synchronizable when $\D$ decreases~\cite{footnote5}.

In fact, for $N \gg 1$ and $m = N - 2\kappa -1$, one can show~\cite{footnote6} that $\D$ scales as $\sim\ln N$ for $n_c = N$, while it is almost a constant for $n_c = 1$.  However, this is achieved by having almost half of the total load on a single giant center node, and, as a result, $\lambda_N/\lambda_2$ scales as $\sim N$ for $n_c = 1$, in sharp contrast to bounded behavior for $n_c = N$, making it difficult for the oscillators to synchronize.


The generality of the correlation between homogeneity and synchronizability can be seen in the rigorous bounds on the ratio $\lambda_N/\lambda_2$ that are valid for \emph{any connected}
network~\cite{footnote6}:
\begin{equation}
\biggl(1-\frac{1}{N}\biggr) \frac{k_{\max}}{k_{\min}} \le \frac{\lambda_N}{\lambda_2} \le (N-1) k_{\max} \ell_{\max}^e {\cal D}_{\max} \D,
\label{eqn:bd}
\end{equation}
where $k_{\min}$ and $k_{\max}$ are the minimum and maximum connectivity, ${\cal D}_{\max}$ is the maximum length of the shortest path between two nodes and $\ell_{\max}^e$ is the maximum of the normalized loads on links (edges), similarly defined as for those on nodes.  While $\lambda_N/\lambda_2$ is larger for heterogeneous networks because $k_{\max}/k_{\min}$ is larger, homogeneous networks have smaller ratios $\lambda_N/\lambda_2$ because $k_{\max}$ and $\ell_{\max}^e$ are smaller.  However, it is the combination of network distances being small (so that ${\cal D}_{\max} \D$ is small) and the distributions of connectivities and loads being homogeneous that makes the network more synchronizable.

It has been suggested often in neuroscience that smaller $\D$ in the nervous system should imply more efficient communication between neurons, and therefore $\D$ might have been minimized (along with other fitness functions) by evolution~\cite{stephan2000, karbowski2001, latora2001}.  The results here, however, suggest that natural selection might have increased homogeneity in the distribution of connections (and hence in the load distribution) at the expense of having larger $\D$, in order to enhance synchronizability.  For example, the neurons in some layers of cortical columns are relatively homogeneous in soma size, which has been visualized using the Nissl stain technique~\cite{kandel}.  Since the cell body size of a neuron is correlated with the size of its dendritic and axonal expansions~\cite{braitenberg}, which in turn correlate with the number of synaptic connections it receives and sends, it appears that neuronal subnetworks within cortical layers have relatively homogeneous connectivity.  The ability of such subnetworks to synchronize facilitates their ability to exhibit bursting behaviors that underlie neural computations.  

In conclusion, we have shown that the common belief that smaller networks have better synchronization property can be misleading for a wide class of networks, such as SF and SW networks.  Using a general framework for synchronization stability of oscillator networks with arbitrary interaction topology, we have established that SF and SW networks will have reduced ability to synchronize as the heterogeneity of their connectivity distribution increases, even though the average network distance between the oscillators becomes smaller.  Increased concentration of load on center nodes, or hubs, appears to be responsible for this behavior.  Our results suggest that in order for oscillators in a network to communicate better, and hence to synchronize more effectively~\cite{izhikevich}, a balance between having small communication distance and uniform load distribution is essential.

\begin{acknowledgments}
TN is supported by DARPA/ONR grant N00014-01-1-0943.  AEM and YCL are supported by AFOSR under Grant No.~F49620-98-1-0400.  FCH is supported in part by NSF grant DMS-0109001.
\end{acknowledgments}


\begin{thebibliography}{32}
\bibitem{footnote1} $\lambda_i \ge 0$ since $L$ is positive semi-definite and $\lambda_1
= 0$ because the rows of $L$ have zero sum.  In addition, if the
network is connected, $\lambda_2 > 0$.
\bibitem{footnote3} We note that a different measure of synchronizability is more appropriate in some particular cases, resulting in a possibly different behavior with respect to $\D$~\cite{pf}.
\bibitem{footnote6} The details will be included in an extended paper.
\end{thebibliography}

\end{document}